\documentclass[lettersize,journal, compsoc]{IEEEtran}
\usepackage{amsmath,amsfonts}
\usepackage{algorithmic}
\usepackage{algorithm}
\usepackage{array}
\usepackage[caption=false,font=normalsize,labelfont=sf,textfont=sf]{subfig}
\usepackage{textcomp}
\usepackage{stfloats}
\usepackage{url}
\usepackage{verbatim}
\usepackage{graphicx}
\usepackage{xcolor}
\usepackage{cite}
\usepackage{multirow}
\usepackage{csquotes}
\usepackage[backref=page,hidelinks]{hyperref}
\usepackage[justification=justified,singlelinecheck=false,font=footnotesize, textfont=sf]{caption}

\begin{document}
\title{VortexViz: \\ Finding Vortex Boundaries by Learning from Particle Trajectories}

\author{Akila de Silva, Nicholas Tee, Omkar Ghanekar, Fahim Hasan Khan, Gregory Dusek, \\ James Davis and Alex Pang
\IEEEcompsocitemizethanks{\IEEEcompsocthanksitem{Akila de Silva is with San Francisco State University. E-mail: \\ desilva@sfsu.edu}}
\IEEEcompsocitemizethanks{\IEEEcompsocthanksitem{Nicholas Tee, Omkar Ghanekar, Fahim Hasan Khan, James Davis, Alex Pang are with UC Santa Cruz. E-mail: audesilv@ucsc.edu, ntee@ucsc.edu, oghaneka@ucsc.edu, fkhan4@ucsc.edu, davis@cs.ucsc.edu, pang@soe.ucsc.edu}}
\IEEEcompsocitemizethanks{\IEEEcompsocthanksitem{Gregory Dusek is with NOAA National Ocean Service. E-mail: \\ gregory.dusek@noaa.gov.}}
\IEEEcompsocitemizethanks{\IEEEcompsocthanksitem{Corresponding authors : Akila de Silva Email: desilva@sfsu.edu and Alex Pang Email: pang@soe.ucsc.edu}}

\thanks{Manuscript received Month XX, 20XX; revised Month XX, 20XX.}
}





\IEEEtitleabstractindextext{%
\begin{abstract}
Vortices are studied in various scientific disciplines, offering insights into fluid flow behavior. 
Visualizing the boundary of vortices is crucial for understanding flow
phenomena and detecting flow irregularities. 
This paper addresses the challenge of accurately extracting vortex boundaries using deep learning techniques. 
While existing methods primarily train on velocity components, we propose a novel approach incorporating particle trajectories ({\em streamlines} or {\em pathlines}) into the learning process. 
By leveraging the regional/local characteristics of the flow field captured by streamlines or pathlines, our methodology aims to enhance the accuracy of vortex boundary extraction.
\end{abstract}

}

\maketitle

\section{Introduction}

Vortices are extensively studied in numerous scientific disciplines to gain insight into the behavior of fluid flows.  In aerodynamics, researchers focus on studying vortices that form in the wake of an aircraft, aiming to mitigate the creation of vortices with long lifetimes; persistent vortices can potentially impede commercial aviation's operational capacity \cite{gerz2002commercial, breitsamter2011wake, rennich1999method}.  Oceanographers, on the other hand, study mesoscale eddies modeled as vortices, to understand the transportation of nutrients and heat in ocean currents  \cite{lguensat2018eddynet,williams2011visualization, wang2021global}.  Additionally, astrophysicists examine the vortical structure in black holes to understand the stability of extremal black holes via topological explanations \cite{blackholes_vortices,lima2023properties,dowker1992euclidean}. Vortices are also studied in high temperature superconductors, to better understand the dissipation of free-current \cite{vortex_superconductors, bardeen1965theory,chapman1995motion}. While the study of vortices transcends numerous scientific disciplines, visualization researchers place particular emphasis on visualizing the vortex boundary to gain insight into fluid flow behavior.

\begin{figure}[t]
    \centering
    \includegraphics[width=0.99\linewidth]{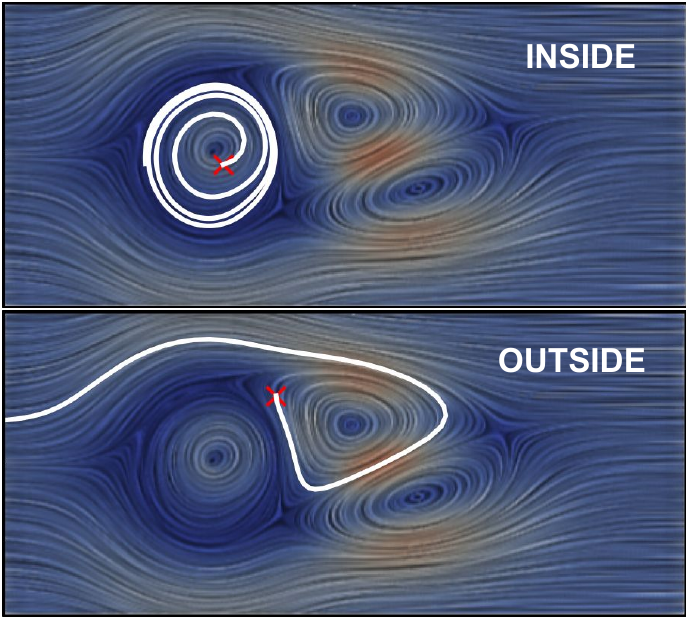}
    \caption{\textbf{Particle Trajectories (pathlines and streamlines) Behave Differently Inside and Outside of a Vortex: } In this paper, we exploit this behavior difference and physical properties of the flow field along the trajectory to find vortex boundaries. (Two streamlines from \cite{Jung93} are shown with a red cross indicating the seed point.) }
    \label{fig:flowlines_intro_figure}
\end{figure}

While there are many definitions of a vortex, it is generally agreed that ``vortices are regions of high vorticity'' with ``multitude of material particles rotating around a common center" \cite{haller2016defining}. Analyzing vortex boundaries is essential for gaining insights into fluid flow behavior, including phenomena such as flow separation, turbulence, and vortex formation and dissipation \cite{jankun2006vortex}. Observing a distorted or irregular shape along the vortex boundary can provide valuable clues about the potential occurrence of vortex breakdown \cite{berenjkoub2020vortex}. While vortex boundary extraction has been studied by visualization researchers for many years \cite{maxworthy1977some}, its precise mathematical definition may vary depending on the specific context \cite{lugt1979dilemma}. In such cases where a formal definition is elusive, deep learning techniques hold great promise in identifying and capturing relevant features.

Supervised deep learning based methods, which rely on labeled datasets, have been  used to extract the vortex boundary \cite{berenjkoub2020vortex, deng2019cnn, deng2022vortex, wang2021rapid, kashir2021application}. These approaches primarily rely on velocity fields represented by their $U$ and $V$ components (velocity components along x and y axes, respectively) to learn about the vortex boundary. However, we contend that learning solely from the velocity components is insufficient in accurately capturing the vortex boundary. This limitation arises from the fact that the velocity field, represented as $U$ and $V$ components fails to effectively capture the non-local behavior of the flow field. To address this issue, we propose an alternative approach in this paper, where we utilize particle trajectories (streamlines or pathlines) instead of velocity fields to learn about the vortex boundary. By incorporating particle trajectories into the learning process, we aim to enhance the model's ability to capture the rotational behavior or the {\em swirliness} of the flow field, thereby improving the accuracy of vortex boundary extraction.


\medskip
\noindent The main contribution of this paper is, 
\begin{itemize}
    \item A novel deep learning methodology utilizing particle trajectories to learn and identify vortex boundaries. 
\end{itemize}
    

\section{Related Work}
\subsection{Deep Learning for Flow Visualization}

\noindent The visualization community has been actively engaged with deep learning in two primary ways: visualization to understand the inner workings of a deep learning models, and use of deep learning in visualization tasks \cite{wang2021survey, survey_dl4flowvis_01, wang2022dl4scivis}.  In the context of the latter, particularly for flow visualization tasks, specific flow features for visualization have been identified with the aid of deep learning.  Some researchers utilized deep learning to find rip currents, a flow pattern found in the near shore ocean \cite{de2021automated, mori2022flow, ripviz}.  Kim and G{\"u}nther \cite{kim2019robust} used a neural network to extract a steady reference frame from an unsteady vector field.  
In \cite{han2022exploratory}, deep neural network based particle tracing method to explore time-varying vector fields represented by Lagrangian flow maps. Numerous studies have also leveraged deep learning to identify vortex boundaries from velocity fields represented as velocity components \cite{berenjkoub2020vortex, deng2019cnn, deng2022vortex, wang2021rapid, kashir2021application}. In this paper, we present an novel deep learning approach that learns to identify vortex boundaries by utilizing information from flowlines (streamlines and pathlines), as opposed to relying solely on velocity components.

\subsection{Threshold-based Methods for Vortex Boundary Detection}

\noindent Threshold-based vortex detection methods can be categorized into two types: local and global methods. Local methods involve the computation of a local quantity at each point within a flow field, resulting in a scalar field. Subsequently, these resulting scalar fields undergo thresholding to identify the contours corresponding to vortices. Most notable local methods are  $Q$ criterion \cite{hunt1987vorticity-Q-criterion}, $\Omega$ criterion \cite{omega_criterion}, $\lambda_2$ criterion \cite{Lambda-2-criterion}, and $\Delta$ criterion \cite{delta-criterion}. However,  these methods may fail to detect obvious vortices while erroneously detecting non-vortical structures. Sadarjoen et al. \cite{winding_angle} argue that vortices are a regional phenomenon, and local methods such as the above criteria are ill-equipped to identify them.

In contrast, global methods such as instantaneous vorticity deviation (IVD) \cite{IVD} and the winding angle method \cite{winding_angle} use global flow information to find vortex boundaries. IVD is defined as the absolute value of the difference between the vorticity at a point in the flow field and the spatially averaged vorticity of the global flow field. There are two main post-processing methods that visualization researchers use to find vortex boundaries using IVD. The first method involves applying a user-defined threshold to the IVD field to detect vortex contours \cite{berenjkoub2020vortex}. However, in certain scenarios, such as when vortices are dissipating, varying thresholds might need to be set for each individual vortex.
On the other hand, other researchers \cite{deng2022vortex, deng2019cnn, wang2021rapid} adopt a different approach by identifying isocontours around vortex cores. Vortex cores are located by identifying local maxima in the IVD field; isocontours satisfying pre-defined arc length and convexity criteria are selected as the vortex boundaries. For this method, the thresholds for local maxima, convexity, and arc length are set by users during runtime. However, the substantial user input required at runtime for these IVD-based approaches makes them less appealing for analyzing large datasets.

The winding angle method, introduced in \cite{winding_angle}, is centered around the identification of streamlines that exhibit rotational behavior around a critical point. Researchers initiate this approach by sparsely seeding streamlines across the flow field. For each streamline, they calculate the sum of signed angles between adjacent line segments. By applying a threshold to this computed sum, streamlines associated with a vortex can be discerned and visually presented to denote vortex locations. It is essential to emphasize that the determination of the threshold value is left to the user and relies on the specific dataset under analysis. Furthermore, it is worth noting that the winding angle method is designed to reveal the locations of vortices by showcasing the streamlines, rather than explicitly outlining the vortex boundary.

\subsection{Deep Learning for Vortex Boundary Extraction}

\noindent In recent years, supervised machine learning methods, dependent on labeled datasets, have been increasingly applied to detect vortex boundaries. Originally developed within the computer vision community to identify specific pixel patterns within images, these methods have been adapted for finding vortices within flowfields.  One notable network architecture frequently employed is U-net \cite{u-net-original}, originally designed for medical image segmentation but adapted to extract vortex boundaries through segmentation of the flow field \cite{berenjkoub2020vortex, deng2022vortex}.
Likewise, various convolutional neural network (CNN) variants have also been tailored to identify vortices in flowfields \cite{deng2019cnn, wang2021rapid,berenjkoub2020vortex}. Furthermore, ResNet has been repurposed to detect vortices within flowfields as well \cite{berenjkoub2020vortex}.


\subsection{Deep Learning Methods with Flowlines}

\noindent The flow visualization community has proposed deep learning methods that learn from flowlines.  Flowlines describe the trajectory of a massless particle of fluid.  In steady state flows (or a snapshot of a time varying flow) they're referred to as streamlines and otherwise referred to as pathlines. Han et al. \cite{han2018flownet} used an autoencoder-based deep learning model, FlowNet, to learn feature representations of streamlines in 3D steady flow fields which are then used to cluster the streamlines. Recently, de Silva et al. \cite{ripviz}, proposed a deep learning LSTM autoencoder methods, that learns about rip currents, a naturally occurring flow pattern in the near shore ocean, by using short sequences of pathlines. In both these works flowlines were encapsulated in binary volumes (for 3D flowlines) or binary images (for 2D flowlines). In this work we use flowlines to find vortex boundaries.

\section{VortexViz}
\begin{figure*}[tb]
    \centering
    \includegraphics[scale=0.5]{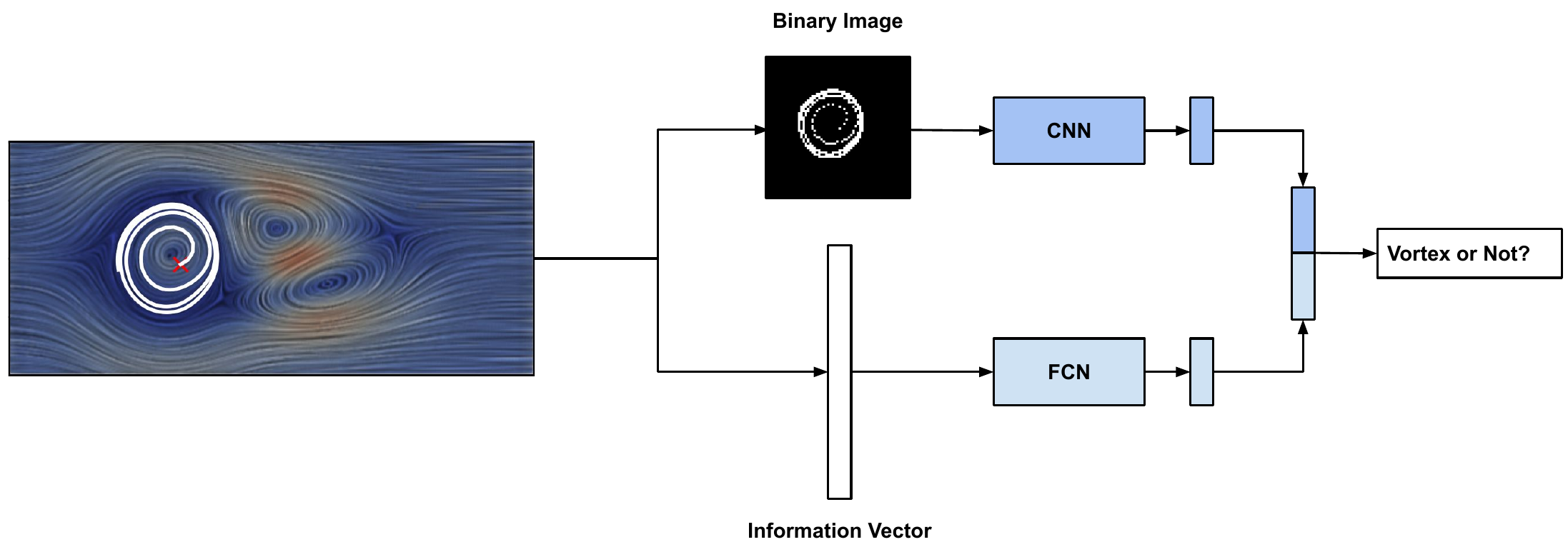}
    \caption{\textbf{VortexViz Pipeline}: Each flowline is represented as a binary image and an information vector. The binary image is processed by a convolutional Neural Network (CNN) and the information vector is processed by a Fully Connected Neural Network (FCN). The intermediate output layers of both these networks are merged. The merged layer is then used to predict if the seed point that originated the flowline is classified as inside a vortex or not. }
    \label{fig:pipeline}
\end{figure*}

\noindent VortexViz identifies points that are within a vortex boundary by combining flowline patterns and information collected along the flowlines.  This is encapsulated in the pipeline shown in Figure \ref{fig:pipeline}\cite{thesis_akila}.  It consists of several steps. Section \ref{sec:flowline_generation} discusses how we generate flowlines. Section \ref{sec:streamline_representation} discusses how we represent flowlines. Section \ref{sec:deep_learning_model} discusses our deep learning model. 

\subsection{Particle Trajectory Generation \label{sec:flowline_generation}}

\noindent We generate particle trajectories from higher-order methods like the widely used fourth-order Runge-Kutta integrator $RK4$ \cite{butcher1996history}. If a particle trajectory travels beyond the domain of the flow field then we stop the numerical integration.  

\begin{figure*}[th]
    \centering
    \includegraphics[width = \textwidth]{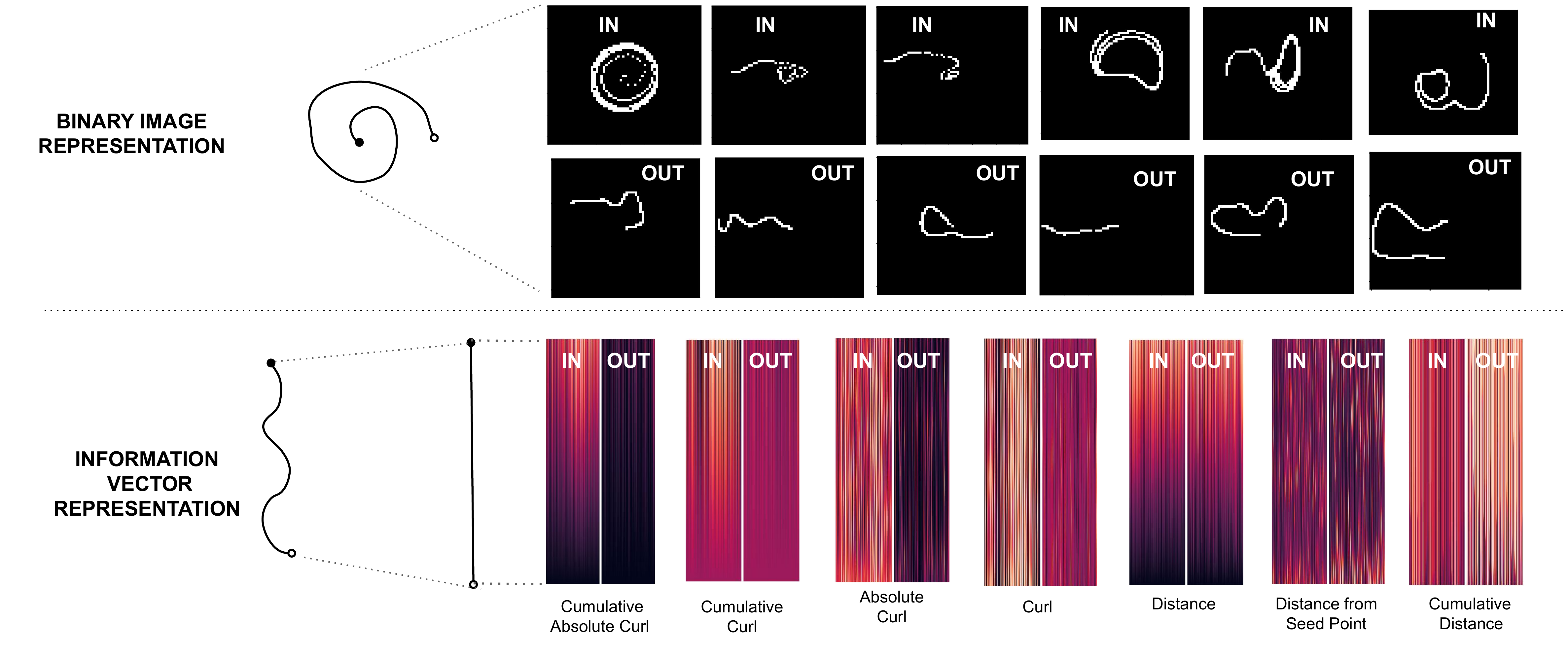}
    \caption{\textbf{Visualizing Binary Images and Information Vectors:} The top half of the figure shows binary image representations of particle trajectories. Notice that binary images exhibit notable disparities for particle trajectories inside (IN) and outside (OUT) of a vortex. The lower half showcases information vector representations of particle trajectories. These information vectors, presented as stacked values in heatmap form, illustrate differences among vectors of equal length. The information vectors initiate from the bottom of the heatmap, aligned with the hollow circle indicating the seed point of the example particle trajectory. Notice that each type of information vector is also different for particle trajectories inside (IN) and outside (OUT) the vortex. These visual differences lead us to believe these representations can be used to detect vortices.}
    \label{fig:binary_image_flowline}
\end{figure*}

\subsection{Particle Trajectory Representation \label{sec:streamline_representation}}

\noindent Each particle trajectory is a collection of ordered points as shown in equation \ref{eqn:flowline} where $(x_i, y_i)$ is a point in the coordinate system of the flow field and $n$ is the maximum number of integration steps of the particle trajectory, while  $(x_1, y_1)$ is the seed point.

\begin{equation}\label{eqn:flowline}
    \mathit{particle\;trajectory}  = \Big \{ (x_1, y_1), \cdots , (x_n, y_n) \Big \}
\end{equation}

In order for the deep learning model to learn about particle trajectories, we need to encode the {\em swirliness} of particle trajectories. To do this, we represent each particle trajectory as a binary image, as discussed in Section \ref{sec:binary_image}, and as an information vector, as discussed in Section \ref{sec:information_vectors}. \\

\subsubsection{\textbf{Binary Image}\label{sec:binary_image}}

\noindent In the binary image, the seed point of each particle trajectory is placed at the center, and the trajectory rescaled to this particle trajectory-centric coordinate system. To do this, each particle trajectory is represented by an
$L \times L$ binary image $\textbf{I}$, where each point in equation \ref{eqn:flowline} is translated  to center the seed point in the binary image. For longer flowlines that extend beyond $L \times L$, we increased the size of the binary image to accommodate these longer flowlines. We then resized all of these larger binary images down to $L \times L$. 
The reason why all flowlines were not simply centered then resized to $L \times L$ is that we need to differentiate flowlines that barely moved from their initial seed position versus those that actually travelled beyond $L \times L$. Each point of the flowline is marked as $1$ in the binary image, while others are marked as $0$. Some examples of  binary images are shown in Figure \ref{fig:binary_image_flowline}.


\subsubsection{\textbf{Information Vector} \label{sec:information_vectors}} 

In addition to using binary images, we hypothesize that physical quantities of the flow field can also be used to capture the {\em swirliness} of the flow field. In particular we observed that information such as curl and distance can be used to differentiate flowlines originating within a vortex from others. We denote these vectors as \textit{information\;vectors} and stored as a $1D$ vector of length $n$, where $n$ is the maximum number of integration steps of the flowline.

We conducted experiments using four different information vectors derived from curl, which gauges the rotational tendency of particles at specific points of the flow field. These are \textit{curl}, \textit{absolute curl} and \textit{cumulative curl}  and \textit{cumulative absolute curl} respectively. The \textit{curl} information vector consists of the curl calculated at each point along the flowline, while \textit{absolute curl} contains the absolute value of the curl at each point of the flowline.  Additionally, \textit{cumulative curl} contains the cumulative value of the curl at each point of the flowline and \textit{cumulative absolute curl} contains cumulative value of the absolute curl at each point of the flowline. Mathematical expressions for information vectors are shown in the supplementary materials.  



In addition to information vectors based on curl, we also explored the use of information vectors based on distance. We noticed that the flowlines seeded in vortices stay a longer time within the domain while the other flowlines exit the domain relatively quickly.
Based on this observation, we encoded two types of information vectors, namely $distance$, where distance between consecutive points of the flowline is used, and \textit{cumulative distance}, cumulative distance travelled upto a point of the flowline is used. Additionally we observed that the due to the swirliness of flowlines seeded in vortices, that the Euclidean distance between the seed point and any other point in the flowline was subjected to a maximum value. Based on this observation, we used \textit{distance from seed point} as another information vector. All mathematical expressions for information vectors are shown in the supplementary materials. 

In order to visualize how these information vectors are different for flowlines seeded in and out of vortices, we visualize the values of stacks of information vectors of the same length as a heatmap as shown in Figure \ref{fig:binary_image_flowline}. Notice that these information vectors inside and outside vortices are visually different, leading us to believe they maybe useful in detecting vortices.




\subsection{Deep Learning Model \label{sec:deep_learning_model}}


The proposed deep learning model learns about each flowline by using two modalities: binary image and information vector as shown in Figure \ref{fig:pipeline}. One branch of the deep learning model learns from the binary image and the other branch learns from the information vector. The binary image is processed through a convolutional neural network (CNN). While the information vector is processed through a fully connected neural network (FCN). The CNNs are used to generate features from $2D$ data, while FCNs are used to generate features from $1D$ data. Then the resulting feature vectors are combined. Finally for each seed point we make the decision if that seed point is in a vortex or not. The loss function used in the method is binary cross entropy. We trained the neural network model using TensorFlow Keras API \cite{chollet2015keras}. The code and the trained models will be provided in the supplementary materials once the paper is accepted for publication.



\subsection{Data Sets}
\noindent We used five unsteady \textit{2D} datasets, namely \textit{2D Unsteady DoubleGyre} \cite{double_gyre}, \textit{2D Unsteady CylinderFlow} \cite{Jung93}, \textit{2D Unsteady Cylinder Flow with von Karman Vortex Street} \cite{vortex_sheet_01, vortex_sheet_2}, \textit{2D Unsteady Beads Problem} \cite{unsteady_beads_01, unsteady_beads_02},  and \textit{2D Unsteady Cylinder Flow Around Corners} \cite{pipedcylinder_01, pipedcylinder_02}. Training data was generated by partitioning  \textit{2D Unsteady DoubleGyre}, \textit{2D Unsteady CylinderFlow} and \textit{2D Unsteady Cylinder Flow with von Karman Vortex Street} into training and test datasets. \textit{2D Unsteady Beads Problem},  and \textit{2D Unsteady Cylinder Flow Around Corners}  were exclusively used for testing. 

While acknowledging the absence of an absolute ``true" definition for a vortex boundary, numerous prior studies \cite{deng2022vortex, berenjkoub2020vortex, wang2021rapid} have commonly utilized the results obtained from IVD (Instantaneous Vorticity Deviation) as a benchmark for comparison. Following the same convention, our study also relied on IVD to establish ground truth for \textit{2D}  datasets, particularly in cases where the IVD output distinctly delineated contours representing vortex boundaries. Notably, within datasets such as \textit{2D Unsteady DoubleGyre}, \textit{2D Unsteady CylinderFlow}, \textit{2D Unsteady Cylinder Flow with von Karman Vortex Street}, we identified unambiguous contours through IVD.

However, for the remaining datasets, namely \textit{2D Unsteady Beads Problem}  and \textit{2D Unsteady Cylinder Flow Around Corners}, we observed that the output derived from IVD did not align with the expected vortex structure as one can perceive from the LIC images. Consequently, we lacked a ground truth for these particular datasets. 

\subsection{Comparison Metric}

\noindent We use \textit{$F_{1}$ score} as the comparison matric as shown in equation \ref{eq:f_1_score}. If a flowline seeded within the vortex boundary is detected as a vortex, then it is counted as a true positive (TPs). Otherwise, it is considered a false positive (FPs). Suppose a flowline originating within the vortex boundary is not detected, then it is a false negative (FNs). The range of the $F_1$ score is between $0$ and $1$. If most flowlines within the vortex boundary are detected as a vortex, the score will be closer to $1$, otherwise closer to $0$.

\begin{equation} \label{eq:f_1_score}
    F_1 = \frac{2}{\frac{1}{recall} + \frac{1}{precision}} = \frac{2}{\frac{FNs+TPs}{TPs} + \frac{FPs+TPs}{TPs}}
\end{equation}

\section{Results and Discussion}
\subsection{Comparison with Existing Methods}

\begin{figure*}
    \centering
    \includegraphics[width=\textwidth]{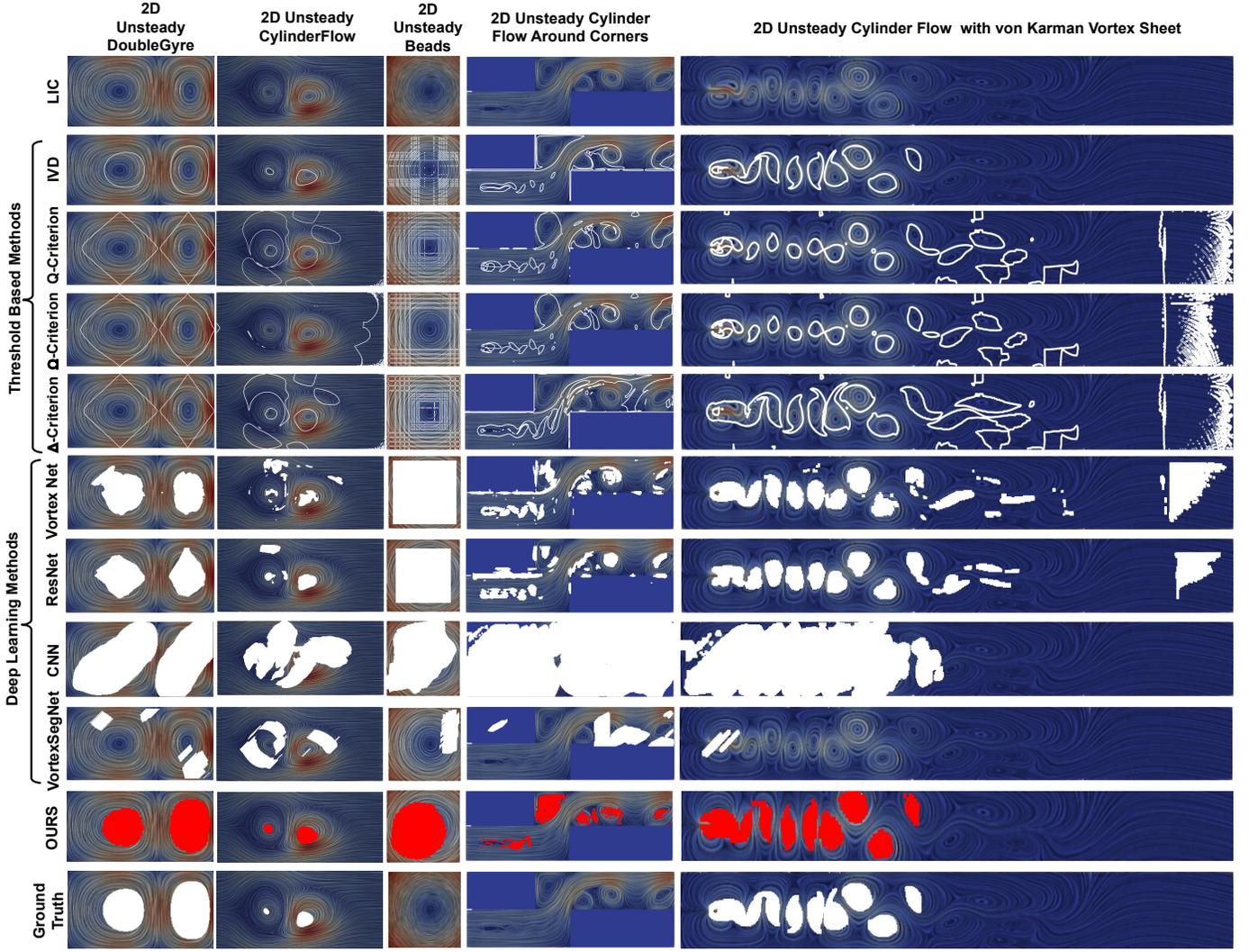}
    \caption{\textbf{Qualitative Comparison with Other Methods.} The first row displays the LIC image. Rows $2$-$5$ exhibit the output of IVD, Q, $\Omega$, and $\Delta$ criteria. Rows $6$-$9$ showcase the output of deep learning methods that learn from velocity components. The second-to-last row presents the output of our method, while the last row depicts the ground truth. Notice that our method can visualize vortices even where other methods fail.} 
    \label{fig:related_work_comparison}
\end{figure*}

\begingroup

\setlength{\tabcolsep}{12pt} 
\renewcommand{\arraystretch}{1.3} 

\begin{table*}
    \centering
    \caption{\textbf{Quantitative Comparison with Other Methods.} We use three datasets where IVD can produce an unambiguous ground truth for quantitative comparisons in this table. Notice that our method has a higher $F_1$ score compared to other methods. For the remaining two datasets IVD did not produce reliable ground truth contours that represent vortices. Therefore we did not include those in this table; however for a qualitative comparison on all datasets please refer to Figure \ref{fig:related_work_comparison}. }
    \begin{tabular}{|c|c|c|c|}
        \hline
          Method   & \shortstack{2D Unsteady \\ DoubleGyre \cite{double_gyre}} & \shortstack{2D Unsteady \\ CylinderFlow \cite{Jung93}} 
          & \shortstack{2D Unsteady \\ Cylinder Flow  \\with \\ von Karman \\Vortex Street \\ \cite{vortex_sheet_01, vortex_sheet_2}}\\
        \hline \hline
       Q criterion  \cite{hunt1987vorticity-Q-criterion}   &  $0.698$  & $0.107$ & $0.311$\\
      \hline
        $\Omega$ criterion \cite{omega_criterion}   &  $0.707$  & $0.124$ & $0.363$\\
      \hline
        $\Delta$ criterion \cite{delta-criterion}   &  $0.657$  & $0.029$ & $0.445$\\
       \hline \hline
       Vortex Net  \cite{deng2019cnn}   &  $0.818$  & $0.332$ & $0.635$\\
      \hline
      ResNet \cite{berenjkoub2020vortex}   &  $0.831$  & $0.550$ & $0.646$\\
      \hline
      CNN \cite{berenjkoub2020vortex}   &  $0.567$  & $0.090$ & $0.360$\\
      \hline
      Vortex Seg Net \cite{wang2021rapid}   &  $0.173$  & $0.008$ & $0.196$\\
        \hline \hline
      Ours   & \textbf{0.972}  & \textbf{0.797} & \textbf{0.946}\\
      \hline
    \end{tabular}
    \label{table_comp_deep_learning}
\end{table*}

\endgroup

\noindent We compare our method against threshold-based methods and deep learning methods that learn from velocity components. Rows 2-5 of Figure \ref{fig:related_work_comparison} shows the results of threshold -based methods and Rows 6-9 of Figure \ref{fig:related_work_comparison} shows the output of the deep learning methods on 5 different datasets. Row 10 shows the output of our method and last row shows the ground truth.

\subsubsection{Comparison with threshold-based methods}

First, we compared our method to threshold-based methods such as \textit{IVD} \cite{IVD}, \textit{$Q$ criterion} \cite{hunt1987vorticity-Q-criterion}, \textit{$\Omega$ criterion} \cite{omega_criterion} and \textit{$\Delta$ criterion} \cite{delta-criterion}. In particular for datasets \textit{2D Unsteady DoubleGyre}, \textit{2D Unsteady CylinderFlow} and \textit{2D Unsteady Cylinder Flow with von Karman Vortex Street} we were able to use \textit{IVD} at dataset specific thresholds that capture the locations of rotational behavior seen in the \textit{LIC} images.. However, for \textit{$Q$,$\Omega$,$\Delta$ criteria}, there was no single threshold that captured the rotational patterns without introducing erroneous contours as well.

Moreover, we found that threshold-based methods could not find contours that represent the vortex boundary for some datasets such as \textit{2D Unsteady Beads Problem}. 
In this particular dataset, the U component (velocity component along the x-axis) is constant across each row; therefore, the partial derivative of the U components along the y-axis, $\partial{u}/\partial{y}$,  is also constant for each row while $\partial{u}/\partial{x}$ is $0$. Likewise, the V component  
 (velocity component along the y-axis) is constant across each column, and therefore the partial derivative of the V component along the x-axis, $\partial{v}/\partial{x}$, is also constant for each column, while $\partial{v}/\partial{y}$ is $0$. Therefore, mathematical expressions that rely on partial derivatives of velocity components, such as curl, will also produce constant values along rows and columns. Consequently, threshold-based methods, which are derived from curl, produce linear (horizontal and vertical) features around vortex cores instead of the usual circular contours.
In comparison, since our method employs flowlines, we can find the vortex. 

Additionally, we found that for \textit{2D Unsteady Cylinder Flow Around Corners}, IVD and $\Delta$ criterion would highlight the boundary of the obstacle corners as vortices. In contrast, our method does not highlight the  corners as a vortex. Visual comparison of the output for each threshold based method is shown in Figure \ref{fig:related_work_comparison} and numerical comparison is given in Table \ref{table_comp_deep_learning}.

\subsubsection{Comparison with deep learning based methods}

Next, we compared our method to existing deep learning approaches that use velocity components such as \textit{VortexNet}  \cite{deng2019cnn}, \textit{ResNet} \cite{berenjkoub2020vortex}, \textit{CNN} \cite{berenjkoub2020vortex} and \textit{VortexSegNet} \cite{wang2021rapid}. In general, we found that these methods would predict vortices even in places of the flow field where there are no vortices present, as shown in Figure \ref{fig:related_work_comparison}. These models learn about vortex formations primarily from velocity fields represented by their velocity components U and V. We hypothesize that learning from velocity components is ineffective in learning the vortex boundary. In velocity components, the defining features—regions of distinctive values—are often distributed away from the vortex core. Understanding the interpretation of vortices by existing deep learning models trained on velocity components is crucial in comprehending the limitations of these methods.

\begin{figure}
    \centering
    \includegraphics[width=0.5\textwidth]{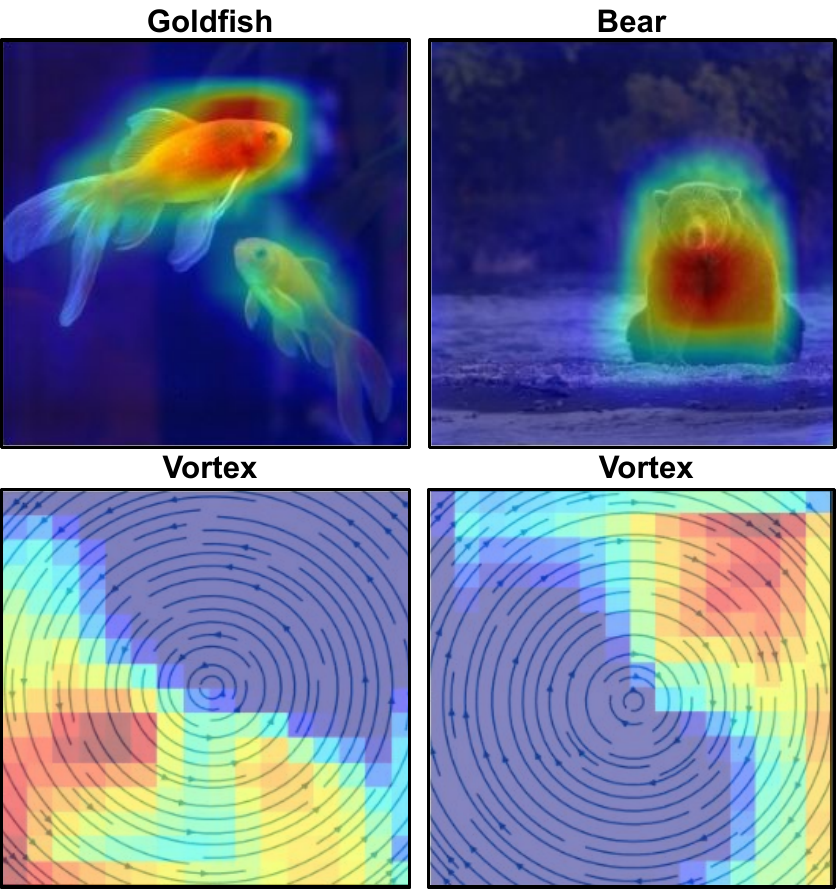}
    \caption{\textbf{Understanding what Deep Learning Methods Learn from Velocity Components using Grad-cam:} In the first row, we display images of a goldfish and a bear. The warmer sections within the overlaid heatmap depict the specific features learned by a CNN. Notably, these features correspond directly to the appearances of the goldfish and bear in the images.
    In the second row, we illustrate a counterclockwise and a clockwise vortex. The warmer regions within the overlaid heatmap reveal the features interpreted by a CNN as indicative of a vortex. It's worth observing that these highlighted features do not align with the vortex core. This observation leads us to hypothesize that what the CNN learns does not necessarily relate to the vortex itself. }
    \label{fig:grad_cam}
\end{figure}

Since most existing deep learning techniques use convolutional layers, we can employ explainable methodologies like Gradient-weighted Class Activation Mapping (Grad-cam) \cite{Grad-cam} to gain insights into what convolutional layers perceive as vortices. Grad-cam identifies parts of the input matrix that most impact the decision of the neural network. In Figure \ref{fig:grad_cam}, we showcase the application of Grad-cam, initially designed for image analysis, to the context of velocity fields. The top row demonstrates how CNNs trained to identify object classes like \textit{goldfish} and \textit{bear} from images learn distinctive features. These features are highlighted by a heatmap, where warmer colors indicating places in the image that contribute more to the decision of the convolutional neural network. Notice that these features correspond directly to the appearances of the goldfish and bear in the images. However, in the second row of Figure \ref{fig:grad_cam}, we show one counterclockwise votex and one clockwise vortex and what a CNN perceives as a vortex using Grad-cam. Notice that the warmer parts of the heatmap is not  aligned with the vortex core. The convolutional operation in the neural network, originally designed to detect edges in images, may not effectively capture the swirling nature of vortices when applied to flow fields. It tends to focus on the prevalent values of the individual components of the flow field rather than capturing the swirliness of vortices. In contrast, our approach differs from this since the model is not learning from the velocity components directly. Instead our model is learning from the pattern and information collected along flowlines.  This representation captures the swirliness found in vortices better than velocity components.

\subsection{Performance Assessment on Noisy Data}


\noindent In our evaluation, we tested the robustness of our method by subjecting the input data to different noise levels, specifically at rates of $1\%$, $5\%$, and $10\%$, by introducing Gaussian noise into the data sets. The primary aim was to gauge and compare the robustness of our method with other methods across varying noise levels.

\begin{figure}
    \centering
    \includegraphics[scale=0.25]{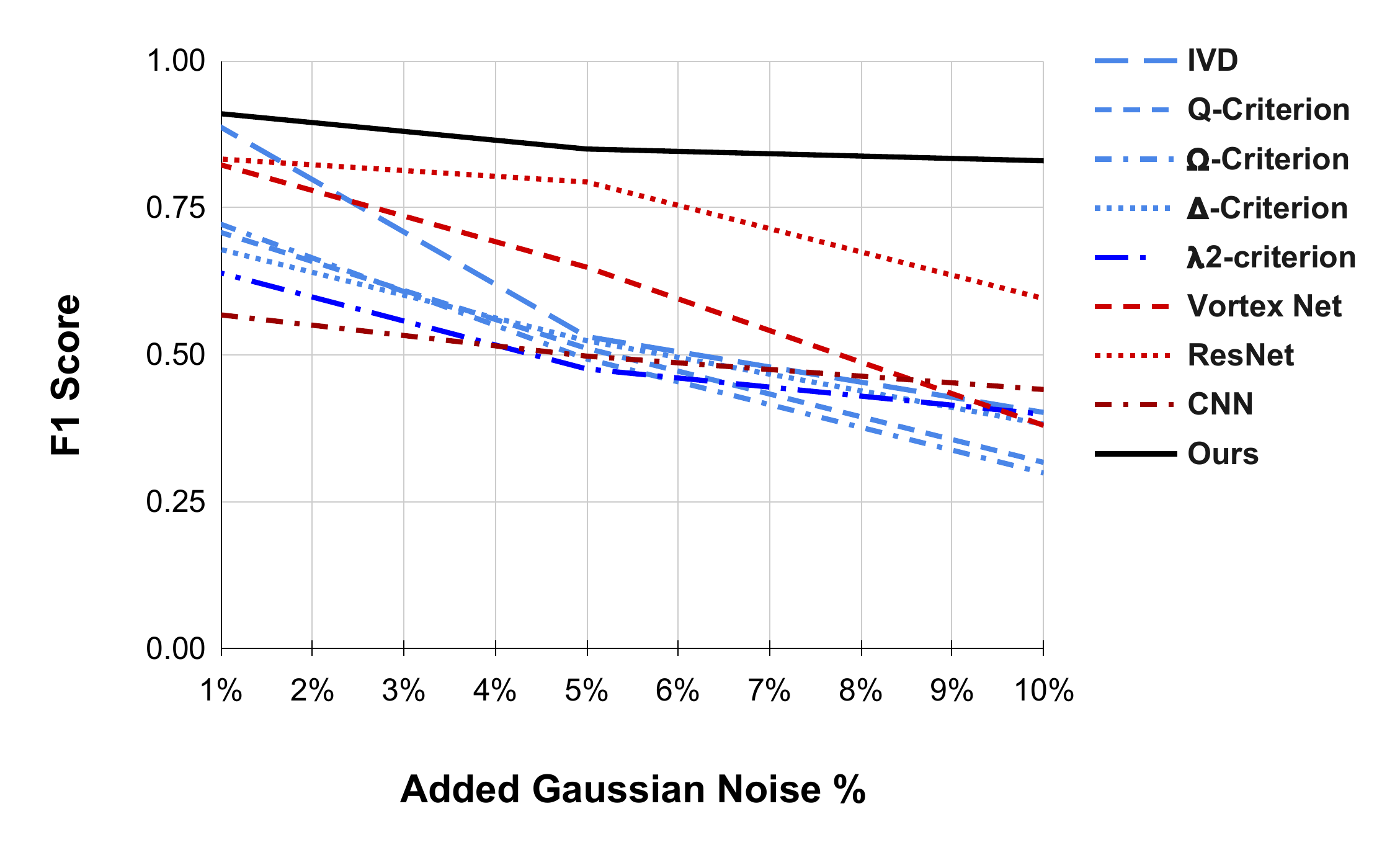}
    \caption{\textbf{Quantitative Comparison of Performance on Noisy Data: } When introducing gaussian noise, we expect a decline in performance across all methods. However, notice that our method (shown in black) exhibits the least decrease in $F_1$ score compared to velocity component based deep learning methods (shown in red) and threshold-based methods (shown in blue).}
    \label{fig:noisy_data_f1_loss}
\end{figure}

\begin{figure*}
    \centering
    \includegraphics[width=\textwidth]{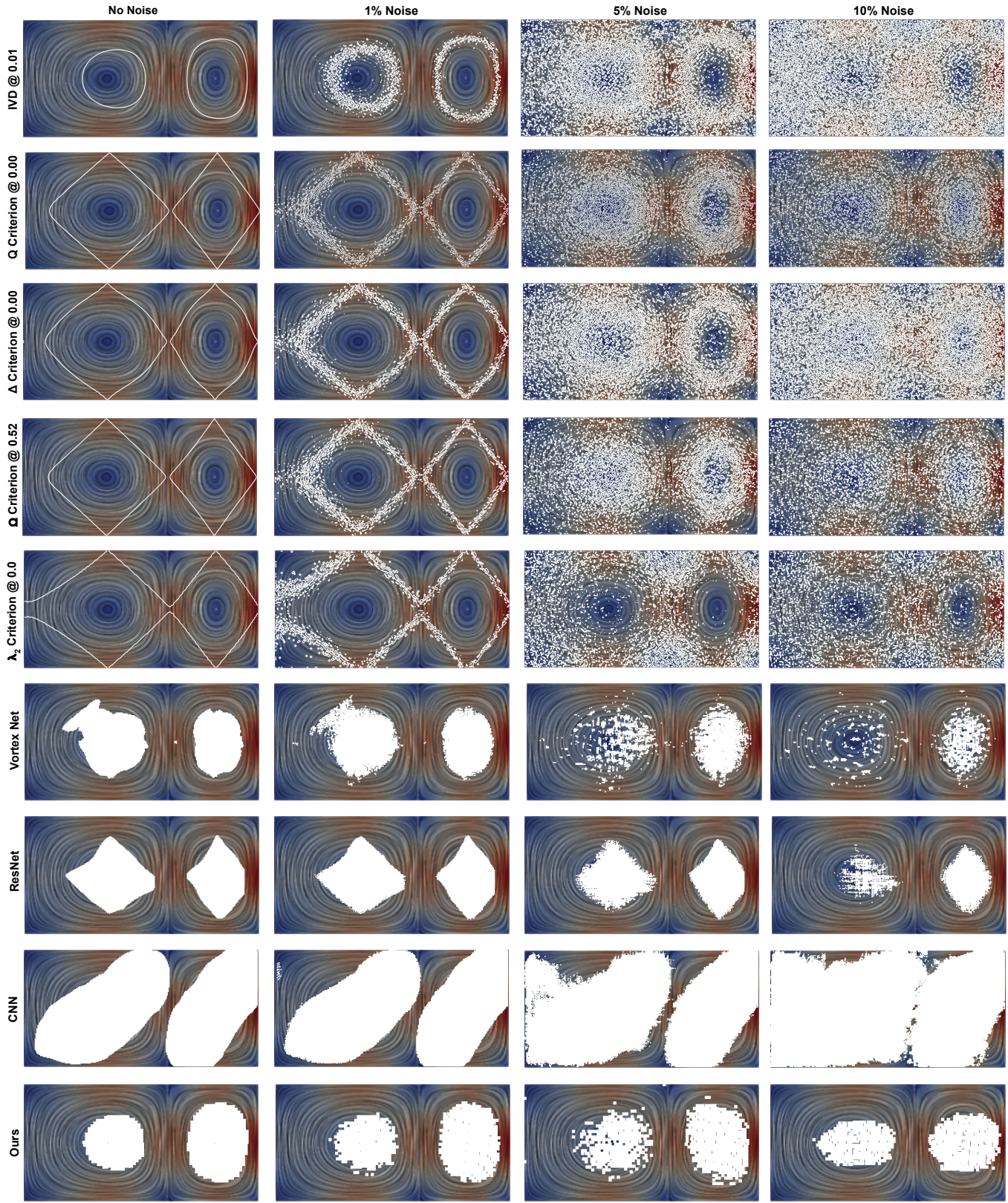}
    \caption{\textbf{Robust with Noisy Data:} In Rows 1-5, the contours based on IVD, Q, $\Delta$, $\Omega$ and $\lambda_2$ criteria are displayed. It is noticeable that as noise levels rise, the discernibility of vortices within these contours diminishes. Similarly, ResNet, VortexNet and CNN (Rows 6-8, respectively) exhibit reduced capability in detecting vortices as noise levels increase. In contrast, our method, depicted in the last row, can maintain the shape of the vortex even amid increased noise levels. } 
    \label{fig:noisy_data_doublegyre}
\end{figure*}

The outcomes, as depicted in Figure \ref{fig:noisy_data_doublegyre}, on the double gyre dataset, highlight the robustness of our method. Even amidst noisy conditions, our approach preserved the circular shape of the vortex. Notably, our observations revealed the sensitivity of threshold-based techniques, which produced deteriorating vortex shapes under a mere 1\% of noise. Conversely, deep learning methods demonstrated robustness, maintaining the vortex's shape with minimal degradation at 1\% noise. However, their $F_1$ scores notably declined when noise level is above 5\%, marking a rapid deterioration in predicting a circular vortex shape as shown in Figure \ref{fig:noisy_data_f1_loss}.

We extended our evaluation to a real-world scenario by utilizing a dataset reconstructed from dense optical flow extracted from a sequence of satellite imagery, specifically selecting the satellite video capturing Hurricane Dorian \cite{noaasatellites_dorian_2019}. Within this context, noise sources included video compression artifacts and errors in deriving the velocity field using optical flow.

\begin{figure*}
    \centering
    \includegraphics[width=\textwidth]{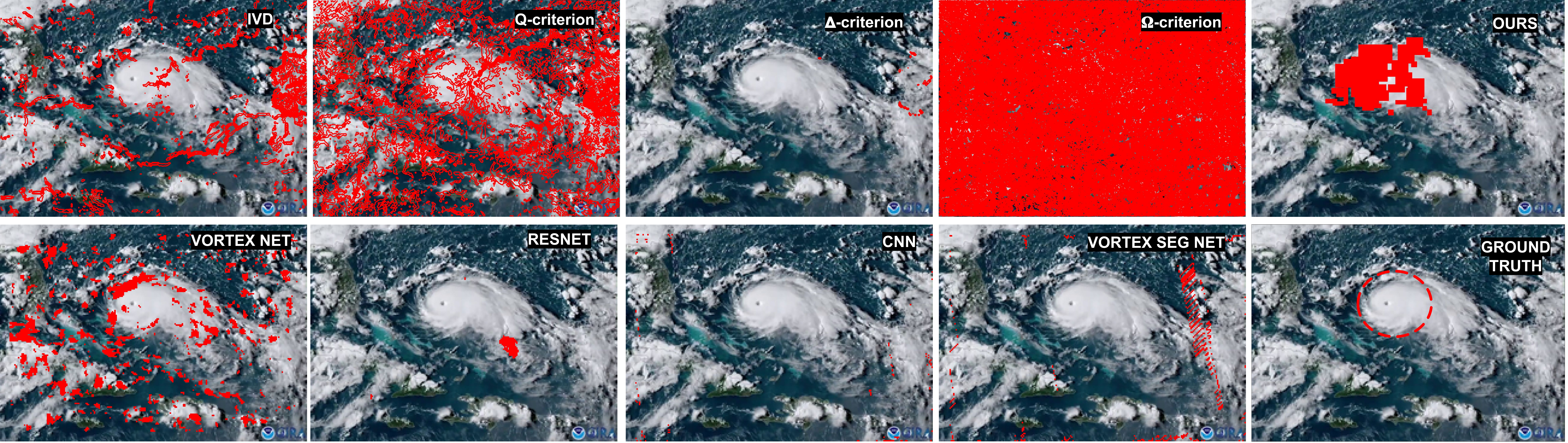}
    \caption{\textbf{Performance Showcase on Real-World Data (Hurricane Dorian, 2019)} Despite inherent noise in the reconstructed velocity field derived from video via optical flow, our method stands out in indicating the approximate location of the vortex amidst the noise. While all methods struggle to identify the vortical structure of the hurricane, our approach excels in delineating the general region of the vortex, even under noisy conditions.}
    \label{fig:hurrican noisy dataset}
\end{figure*}

None of the methods, including ours, was optimized for a hurricane dataset. However, we discovered that our method can visualize the general area of the hurricane vortex, further underscoring its robustness under noisy conditions, as shown in Figure \ref{fig:hurrican noisy dataset}

While the performance of our method on this dataset was surprising, we recognize the need for further optimization to enhance our method's precision in detecting the hurricane vortex. This outcome serves as a promising starting point, prompting us to refine our approach for more accurate extraction of vortex boundaries in real-world scenarios.

\subsection{Sensitivity Analysis of VortexViz Components}

\subsubsection{Pairing of Information Vectors with Binary Images}

\noindent We conducted experiments to determine the optimal pairing of information vectors and binary images, outlined in Table \ref{tab:info_vector_analysis}. It became evident that relying solely on the shape of flowlines encapsulated within binary images did not yield satisfactory outcomes. While using flowline shape within binary volumes \cite{han2018flownet} and binary images \cite{ripviz} has proven successful in various flow visualization tasks, for precise vortex boundary detection, depending solely on binary images, proved inadequate.

Moreover, our experiments involved pairing the binary image with different information vectors. The initial set of information vectors is derived from flowline point distances. Surprisingly, all distance-based information vectors—namely, \textit{Distance}, \textit{Cumulative Distance}, and \textit{Distance from seed point}—both in isolation and when paired with a binary image, failed to perform as well as other parings. 

The second set of information vectors were derived from curl such as \textit{Curl}, \textit{Absolute Curl}, \textit{Cumulative Curl} and \textit{Cumulative Absolute Curl}. We noticed that these information vectors had superior performance when used on their own and when paired with a binary image. We attribute this performance superiority to the fact that curl along the flowline can capture the rotational behavior of a vortex. Remarkably, apart from \textit{Curl} alone, these vectors maintained robustness even under noisy conditions. Notably, the combination of the binary image with \textit{Cumulative Absolute Curl} emerged as the most effective under these circumstances. This paring was used for the results presented in this paper. Our approach quantifies the general consensus regarding vortices as \textit{concentrated regions of high vorticity} \cite{haller2016defining}.

\begingroup

\setlength{\tabcolsep}{5pt} 
\renewcommand{\arraystretch}{1.3} 

\begin{table}
    \centering
    \caption{\textbf{Sensitivity Analysis of Combinations of Information Vectors and Binary Images.} Notice that \textit{binary image + cumulative absolute curl} has the hightest $F_1$ score especially in noisy data.}
    \begin{tabular}{|c|c|c|}
        \hline
          Method   & 0\% noise & 10\% noise\\
        \hline \hline
      Binary Image \textit{Only}    &  $0.814$  & $0.726$ \\
      \hline
      Distance \textit{Only}   &  $0.034$  & $0.004$\\
      \hline
      Cumulative Distance \textit{Only}   &  $0.063$  & $0.004$\\
      \hline
      Distance from Seed Point \textit{Only}    &  $0.660$  & $0.263$\\
        \hline 
      Curl \textit{Only}   & 0.947  & 0.254\\
      \hline
            Absolute Curl \textit{Only}   & 0.952  & 0.737\\
      \hline
            Cumulative Curl \textit{Only}   & 0.961  & 0.721\\
      \hline
            Cumulative Absolute Curl \textit{Only}   & \textbf{0.976}  & 0.749\\
      \hline \hline
            Binary Image + Distance   & 0.801  & 0.420\\
      \hline
            Binary Image + Distance from seed point   & 0.822  & 0.510\\
      \hline 
           Binary Image + Cumulative Distance   & 0.799  & 0.718\\
      \hline 
            Binary Image + Curl    & 0.962  & 0.686\\
      \hline 
            Binary Image + Absolute Curl   & 0.954  & 0.711\\
      \hline 
            Binary Image + Cumulative Curl    & 0.969  & 0.731\\
      \hline 
            Binary Image + Cumulative Absolute Curl    & \textbf{0.972}  & \textbf{0.781}\\
      \hline 
      
    \end{tabular}
    \label{tab:info_vector_analysis}
\end{table}

\endgroup

\subsubsection{Choosing Between Pathlines and Streamlines}

\noindent We conducted experiments to find the optimal type of flowline for our method. Flowlines track the trajectory of a massless fluid particle, termed as streamlines in steady-state flows or in snapshots of time-varying flows, and as pathlines in other cases. In our investigation detailed in Table \ref{tab:pathline_vs_streamline}, we explored both pathlines and streamlines with our approach. While streamlines performed consistently well, the effectiveness of pathlines varied depending on the dataset.

We found both streamlines and pathlines effective when the vortex cores remain fairly stationary
such as in the \textit{2D Unsteady DoubleGyre} , \textit{2D Unsteady CylinderFlow} or \textit{2D Unsteady Beads Problem}. However, in instances where vortices exhibit translational movement over time, as observed in the \textit{2D Unsteady Cylinder Flow with von Karman Vortex Street} or \textit{2D Unsteady Cylinder Flow Around Corners}, streamlines exclusively demonstrated effectiveness, whereas pathlines did not yield favorable results. Our findings suggest that the suitability of flowline type depends upon the dynamic behavior of vortices, with streamlines exhibiting more consistent performance across various scenarios compared to pathlines.

\begingroup

\setlength{\tabcolsep}{7pt} 
\renewcommand{\arraystretch}{1.4} 

\begin{table}
    \centering
    \caption{\textbf{Pathlines or Streamlines: } While streamlines performed consistently higher $F_1$ score, the effectiveness of pathlines varied depending on the dataset.}
    \begin{tabular}{|c|c|c|}
        \hline
          Dataset   & Streamlines & Pathlines\\
        \hline \hline
      2D Unsteady DoubleGyre \cite{double_gyre} &  $\textbf{0.972}$ & $\textbf{0.972}$ \\
      \hline
      2D Unsteady CylinderFlow \cite{Jung93}    & $\textbf{0.797}$ & $\textbf{0.795}$\\
      \hline
      \multirow{2}{*}{\shortstack{2D Unsteady Cylinder Flow with \\ von Karman Vortex Street \cite{vortex_sheet_01, vortex_sheet_2}}} & \multirow{2}{*}{$\textbf{0.946}$} & \multirow{2}{*}{$0.391$}\\
      & &  \\
      \hline
      
    \end{tabular}
    \label{tab:pathline_vs_streamline}
\end{table}

\endgroup

\subsubsection{Optimal Flowline Length and Binary Image Size}
\begin{figure}
    \centering
    \includegraphics[width=0.5\textwidth]{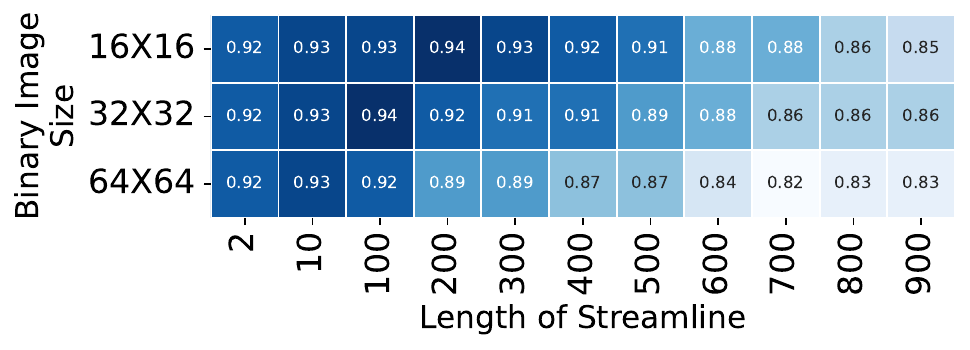}
    \caption{\textbf{Exploring the Impact of Varying Flowline Lengths and Binary Image Sizes on Binary Image + Cumulative Absolute Curl Pairing}. Notice that longer flowlines tend to perform poorly and scaling up binary image sizes did not significantly enhance the $F_1$ score. }
    \label{fig:flow_line_length_vs_binary_image_size}
\end{figure}

Our experiments focused on determining our method's ideal flowline length and binary image size. Figure \ref{fig:flow_line_length_vs_binary_image_size} illustrates a notable trend: longer flowlines showed a decline in accuracy. We attribute this decrease to flowlines exiting the domain before completing integration over an extended interval. Additionally, our observations indicated that increasing binary image sizes did not notably improve the $F_1$ score. For the results presented in this work, we used binary images of size $16 \times 16$ and a flowline of length $200$.

\subsubsection{Comparison of Numerical Integration Methods for Flowline Generation}

Traditionally, the flow visualization community favors higher order numerical integration methods for their superior accuracy \cite{laramee2004state}. However, in our experiments, we examined both higher order (specifically, fourth order Runge Kutta) and lower order (such as first order Euler) integration methods to explore their impact on deep learning's understanding from flowlines.

Higher-order methods like the widely used fourth order Runge Kutta integrator employ more function evaluations per step to better capture the local behavior of integral curves. Conversely, lower order methods like Euler integration are more straightforward but less accurate in capturing intricate flow behaviors.

Despite the community preference for higher order integration methods in flow visualization, our findings revealed that utilizing lower order integration methods for our machine learning model did not notably compromise its performance. This unexpected observation suggests that while higher order methods excel in flow visualization tasks, lower order methods can still effectively contribute to machine learning-based analysis of flowlines without significantly impairing model performance. We followed the norm of using RK4 but note that one can use Euler if limited computational resource is a hindrance.

\begin{figure}
    \centering
    \includegraphics[width=\linewidth]{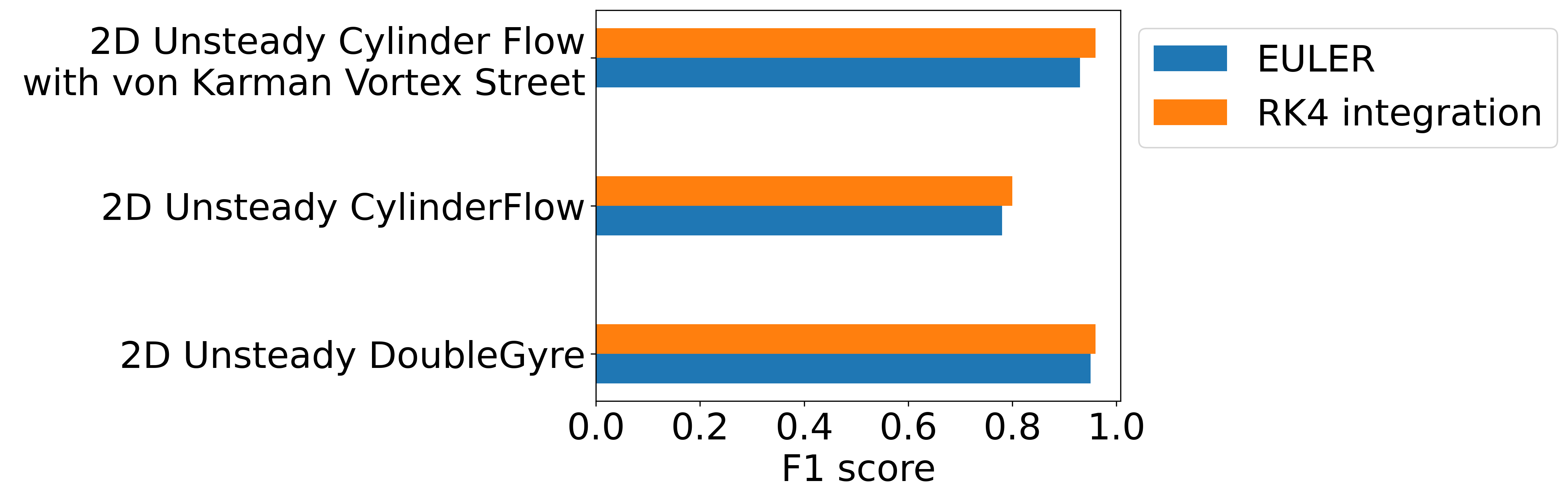}
    \caption{\textbf{Higher Order or Lower Order Numerical Integration Methods: } Notice that there is no significant improvement in using higher order numerical integration methods }
\end{figure}

\section{Conclusion}
\noindent The visualization community has been using deep learning methods directly on velocity components to find vortices. In this paper, we present a novel deep learning approach that learns from flowlines to find vortex boundary. The main contribution of this paper is a deep learning methodology utilizing flowlines to learn and identify vortex boundaries.

Hyperparameters of VortexViz were found based on the datasets tested in this paper.
It is possible that further adjustments may be needed if a new dataset is markedly different from those used in this paper.

For each comparison paper, we used our best judgement
to infer and reproduce the deep learning models associated with each proposed method.
Our inference is based on the technical information provided by 
each paper. For most of the comparison papers the complete code, trained model weights, test and training data are not publicly available. Therefore, we do not claim that we have perfectly captured the authors’ intentions. In order to make it easier for other researchers to improve upon our findings, the code for VortexViz will be available in the supplementary materials after the paper is accepted for publication. 

\section*{Acknowledgments}
This report was prepared in part as a result of work sponsored by the Southeast Coastal Ocean Observing Regional Association (SECOORA) with NOAA financial assistance award numbers NA20NOS0120220, NA23NOS0120243. The statements, findings, conclusions, and recommendations are those of the author(s) and do not necessarily reflect the views of SECOORA or NOAA. 

This project is funded, in part, by the US Coastal Research Program (USCRP) as administered by the US Army Corps of Engineers® (USACE), Department of Defense. The content of the information provided in this publication does not necessarily reflect the position or the policy of the government, and no official endorsement should be inferred.

Partial support for some authors was provided by WISEautomotive through the ATC+ Program award from Korean MOTIE and the Center for Coastal Climate Resilience.


\bibliographystyle{IEEEtran} 
\bibliography{template}

\section{Biography Section}
\begin{IEEEbiography}[{\includegraphics[width=1in,height=1.25in,clip,keepaspectratio]{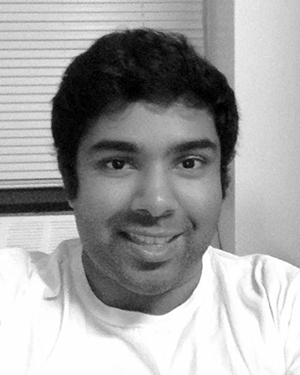}}]{Akila de Silva}
is a assistant professor at San Francisco State University. He has received the Ph.D. in computer science at University of California, Santa Cruz. He has received the M.S. degree from Columbia University in the City of New York and the B.S. degree from the Asian Institute of Technology, Thailand. He is interested in visualization, machine learning, deep learning, and computer vision. 
\end{IEEEbiography}

\begin{IEEEbiography}[{\includegraphics[width=1in,height=1.25in,clip,keepaspectratio]{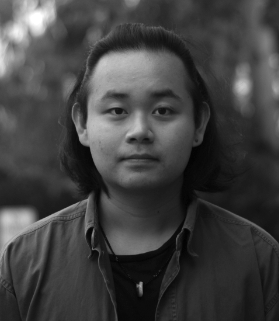}}]{Nicholas Tee} is a M.S. student in computer science at the University of California, Santa Cruz. He has also received his B.S. degree in computer science and his B.A. degree in mathematics there. He is interested in data science, deep learning, and machine learning
\end{IEEEbiography}

\begin{IEEEbiography}[{\includegraphics[width=1in,height=1.25in,clip,keepaspectratio]{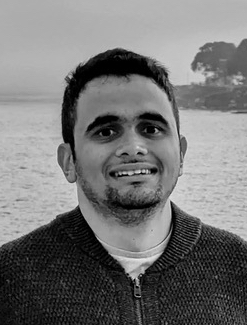}}]{Omkar Ghanekar} is a masters student in the CSE department at UCSC. He has completed his bachelors degree from PICT, India and started his MS in Fall 2021. He is interested in computer vision, visualization. 
\end{IEEEbiography}

\begin{IEEEbiography}[{\includegraphics[width=1in,height=1.25in,clip,keepaspectratio]{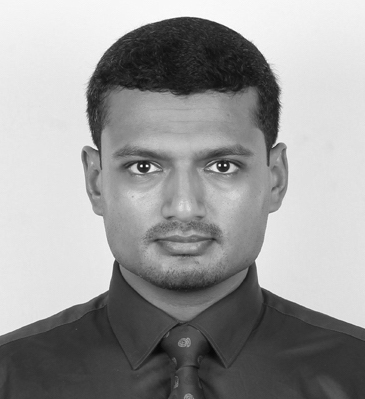}}]{Fahim Hasan Khan} received his B.S. degree from RUET, Bangladesh, and his M.S. degree from the University of Calgary, Canada. He is currently pursuing a Ph.D. degree in computer science at the University of California, Santa Cruz. His research interests include computer graphics, computer vision, applied machine learning, and flow visualization.
\end{IEEEbiography}

\begin{IEEEbiography}[{\includegraphics[width=1in,height=1.25in,clip,keepaspectratio]{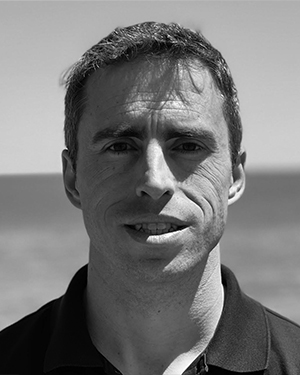}}]{Gregory Dusek}
received the B.S. degree in applied mathematics and the M.S. degree in
teaching and curriculum from the University of Rochester, Rochester, NY, USA, in 2004 and 2005,
respectively, and the Ph.D. degree in marine science from the University of North Carolina at
Chapel Hill, Chapel Hill, NC, USA, in 2011. Since 2014, he has been the Chief Scientist with the
NOAA National Ocean Service, Center for Operational Oceanographic Products and Services. He is currently a Physical Oceanographer.
\end{IEEEbiography}

\begin{IEEEbiography}[{\includegraphics[width=1in,height=1.25in,clip,keepaspectratio]{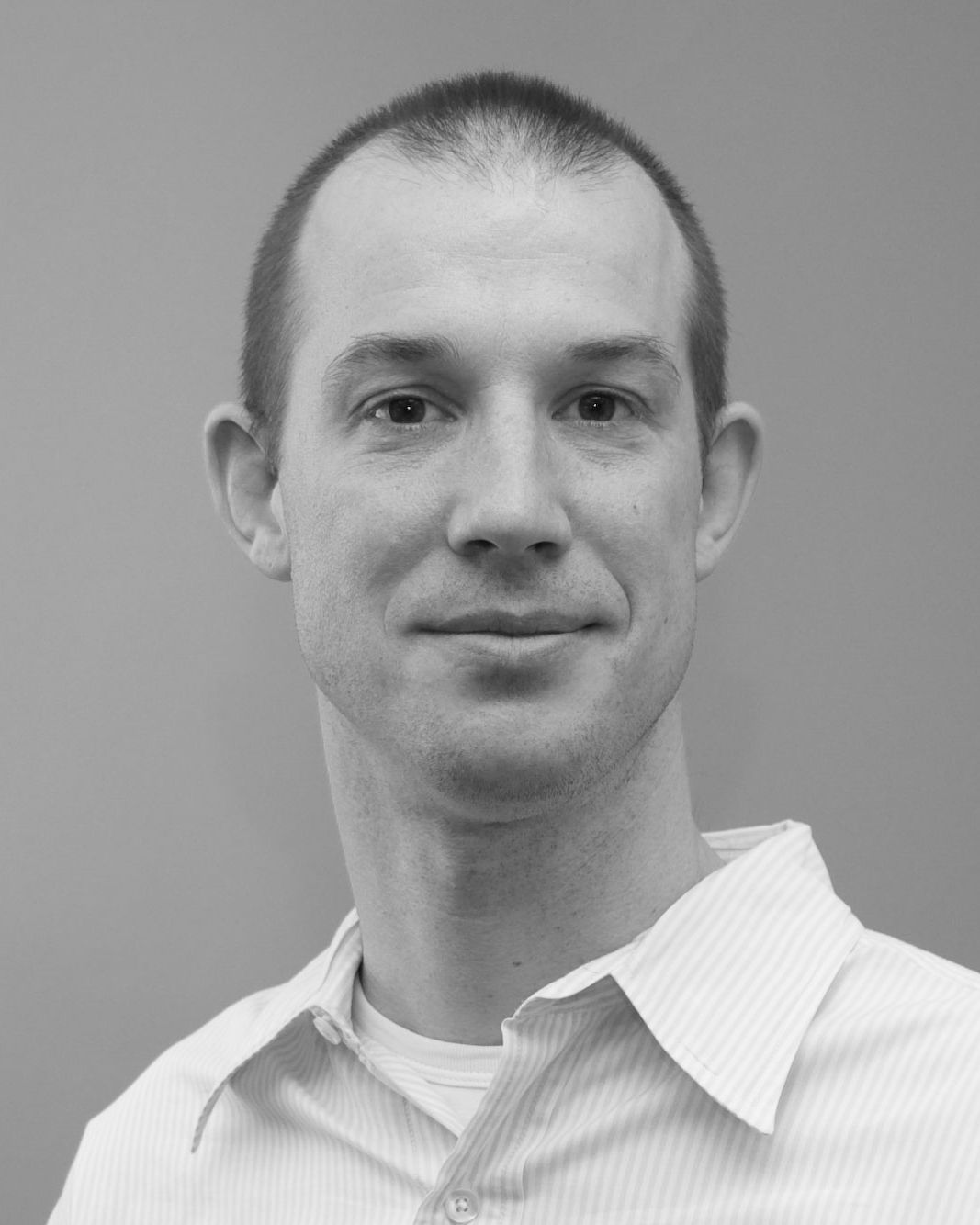}}]{James Davis}
received the Ph.D. degree from Stanford University,
in 2002. From 2002 to 2004, he was a Senior Research Scientist at Honda Research Institute.
He is currently a Professor of computer science and engineering at the University of California at
Santa Cruz. His research interests include both traditional computer science and technology applications to social issues. This work has resulted in over 100 peer-reviewed publications, patents,
and invited talks, received an IEEE ICRA 2003 Best Vision Paper, IEEE
ICCV 2009 Marr Prize, and an NSF CAREER Award. His teaching has twice
been awarded for innovative style, including a course on the importance of
technology to social entrepreneurship. He is on the advisory boards of several
for-profit and non-profit organizations, and co-founded Bellus3D, in 2015.
\end{IEEEbiography}

\begin{IEEEbiography}[{\includegraphics[width=1in,height=1.25in,clip,keepaspectratio]{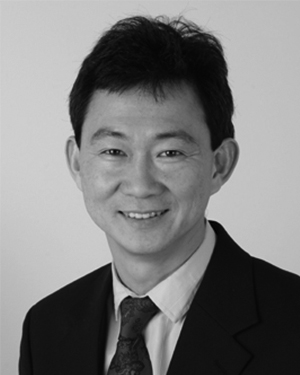}}]{Alex Pang}
is a professor in Computer Science and Engineering
at the University of California Santa Cruz.
He received his PhD in Computer Science from UCLA in 1990,
and his BS in Industrial Engineering from University of the
Philippines with magna cum laude in 1981.
Professor Pang has published over 100 refereed papers in journals and
conferences in both scientific and information visualization.
His research interests include comparative and uncertainty visualization,
flow and tensor visulization, and visualization for social good. 
\end{IEEEbiography}

\clearpage
\newpage

\section{Supplementary Material}
\subsection{Mathematical Expressions of Information Vectors}

The first set of information vectors are derived from curl at each point of the flowline. For each point in the flowline, we calculated the curl as shown in equation \ref{eqn:curl} where $u$ represents the velocity component along the $x$ axis and $v$ represents the velocity component along the $y$ axis. $curl$ measures the tendency at a given point for particles to rotate.

\begin{equation}\label{eqn:curl}
    curl = \Bigg \{ \frac{\partial{v}}{\partial{x}}  - \frac{\partial{u}}{\partial{y}}\biggr\rvert_{\substack{x=x_1\\y=y_1}}, \cdots ,  \frac{\partial{v}}{\partial{x}}  - \frac{\partial{u}}{\partial{y}}\biggr\rvert_{\substack{x=x_n\\y=y_n}} \Bigg \}
\end{equation}

From $curl$, we derive additional information vectors. As shown in equation \ref{eqn:abs_curl} we calculate the absolute value of curl at each point of the flowline. $absolute\;curl$ allows us to encode rotation at each point of the flowline without encoding direction. 

\begin{equation}\label{eqn:abs_curl}
    \mathit{absolute\;curl}  = \Big \{ |curl[1]|, \cdots ,|curl[n]| \Big \}
\end{equation}

In addition to \textit{curl} and \textit{absolute curl}, where physical quantities are calculated at each point of the flowline, we also calculate the cumulative quantities along the flowline. As shown in \ref{eqn:cumu_curl} and \ref{eqn:cumu_abs_curl} we calculate the \textit{cumulative curl}  and \textit{cummulative absolute curl} respectively. We hypothesize that flowlines originating from within vortices have higher \textit{cumulative curl} and \textit{cumulative absolute curl} than flowlines originating from laminar flow regions.

\begin{equation}\label{eqn:cumu_curl}
     \begin{array}{l}
      cumulative\\
      curl
    \end{array}  =  \Bigg \{ \sum_{i=1}^{1}{curl[i]}, \cdots, \sum_{i=1}^{n}{curl[i]} \Bigg \}
\end{equation}

\begin{equation}\label{eqn:cumu_abs_curl}
\begin{split}
    \begin{array}{l}
      cumulative\\
      absolute \\
      curl
    \end{array} 
    =  \Bigg \{ \sum_{i=1}^{1}{\mid curl[i] \mid}, \cdots, \sum_{i=1}^{n}{\mid curl[i] \mid} \Bigg \}
\end{split}
\end{equation}

In addition to the vectors derived from $curl$, we also derived information vectors from the Euclidean distance between successive points of the flowline as shown in equation \ref{eqn:distance}. We hypothesize that the distance traveled by a particle traced from inside a vortex would be different from a particle outside the vortex.

\begin{equation} \label{eqn:distance}
\begin{split}
       distance =  \Big \{ 0, \lVert (x_1, y_1), (x_2, y_2) \rVert , \cdots , \\ \lVert (x_{n-1}, y_{n-1}), (x_n, y_n) \rVert \Big \}
\end{split}
\end{equation}

We also calculated the Euclidean distance between each point of the flowline from the seed point $(x_{1}, y_{1})$ as shown in equation \ref{eqn:seed_point_distance}. We supposed that the distance from the seed point would not exceed some maximum value for particles traced in the vortex.

\begin{equation}\label{eqn:seed_point_distance}
\begin{split}
    \begin{array}{l}
      distance\;from\\
      seed\;point
    \end{array} =  \Big \{ 0, \lVert (x_1, y_1), (x_2, y_2) \rVert , \cdots , \\ \lVert (x_{1}, y_{1}), (x_n, y_n) \rVert \Big \}
\end{split}
\end{equation}

With equation \ref{eqn:distance}, we calculate the $cumulative\;distance$ up to each point in the flowline as shown in equation \ref{eqn:cumulative_distance}

\begin{equation} \label{eqn:cumulative_distance}
\begin{split}
    \begin{array}{l}
      cumulative\\
      distance
    \end{array} =    
      \big \{\sum_{i=1}^{1}{distance[i]}, \cdots,\\
       \sum_{i=1}^{n}{distance[i]} \big \}
\end{split}
\end{equation}

\subsection{Training and Testing Data}

We will provide a link to download training data, test data and code after publication

\subsection{More Grad-cam Images}

\begin{figure}[ht]
    \centering
    \includegraphics[scale=0.5]{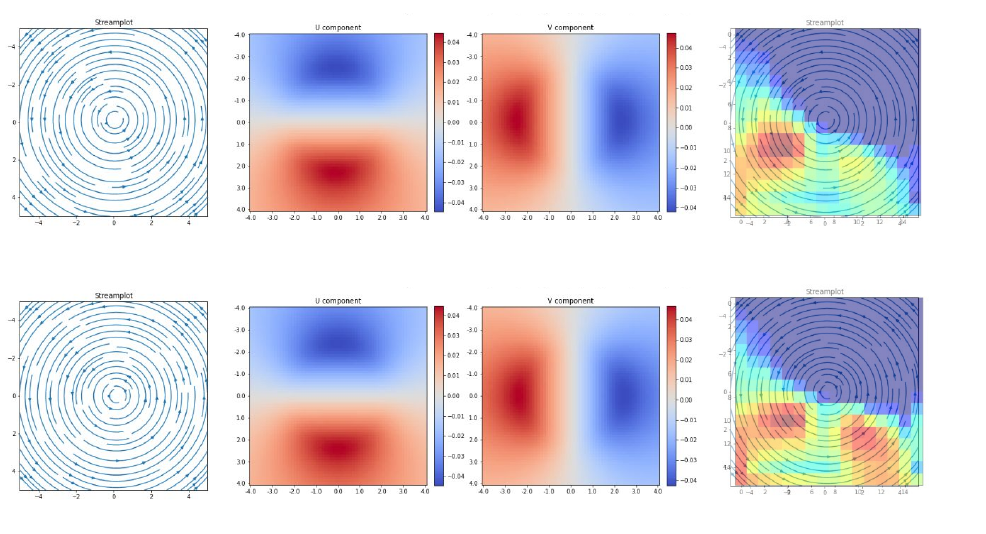}
    \caption{\textbf{Explaining what a CNN learns using Grad-cam:} We illustrate two vortices. The warmer regions within the overlaid heatmap reveal the features interpreted by a CNN as indicative of a vortex. It's worth observing that these highlighted features do not align with the vortex core. This observation leads us to hypothesize that what the CNN learns does not necessarily relate to the vortex itself but learns where higher values (red) are concentrated in the velocity components (two center columns).  }
\end{figure}

\vfill

\end{document}